\documentclass[dvipdfmx]{apex}
\usepackage{txfonts}
\usepackage{graphicx}

\title{Simultaneous control of thermoelectric properties in p-type and n-type materials by electric double-layer gating : New design for thermoelectric device}

\author{Ryohei Takayanagi$^{1}$, Takenori Fujii$^{2}$\thanks{E-mail address: fujii@crc.u-tokyo.ac.jp}, and Atsushi Asamitsu$^{1,2}$}
\inst{$^{1}$Department of Applied Physics, University of Tokyo, Bunkyo, Tokyo 113-8656, Japan \\
$^{2}$Cryogenic Research Center, University of Tokyo, Bunkyo, Tokyo 113-0032, Japan}
\abst{We report novel design for thermoelectric device which can control thermoelectric properties of p-type and n-type materials simultaneously by electric double-layer gating. Here, p-type Cu$_2$O and n-type ZnO were used as positive and negative electrodes of the electric double-layer capacitor structure. When the gate voltage was applied between two electrodes, the holes and electrons were accumulated on the surface of Cu$_2$O and ZnO, respectively. The thermopower was measured by applying thermal gradient along the accumulated layer on the electrodes. We demonstrate here that the accumulated layers are worked as a p-n pair of the thermoelectric device.}

\begin{document}
\maketitle

Since thermoelectric devices can convert waste heat into useful electricity, it is expected to be used as the clean energy without discharging CO$_2$. And also, it can be used as refrigerators through the Peltier effect. Thermoelectric devices are made from an array of p-type and n-type materials connected electrically in series but thermally in parallel. It is therefore important to explore both p-type and n-type materials which have high efficiency in order to make high performance devices.

The efficiency of the thermoelectric materials are determined by the figure of merit of the materials, $ZT =  S^2T / \rho \kappa$, where $T$ is the absolute temperature, $S$ is the Seebeck coefficient, $\rho$ is the resistivity and $\kappa$ is the thermal conductivity. Therefore, the thermoelectric materials needs to have a large $S$, a small $\rho$ and a small $\kappa$. Since $\rho$, $S$, and $\kappa$ are functions of carrier concentration, it is necessary to control and optimize the carrier concentration in order to enhance the figure of merit. Good thermoelectric materials are typically heavily doped semiconductors with carrier concentration of 10$^{19}$ to 10$^{21}$ carriers / cm$^3$\cite{mahan}. 

Most popular way to control carrier concentration is chemical doping, such as chemical substitution or intercalation\cite{ohtaki,BSCSO}. However, it is difficult to synthesize a variety of samples whose doping ratio is precisely controlled. Moreover, the doped elements inherently act as disorder and may lead to the increase of resistivity. The other way to control carrier concentration is a method to use electric field effect\cite{FET1,FET2,FET3}, which is completely free from disorder. It is reported in SrTiO$_3$ that the thermopower was modulated by using field effect transistor (FET) structure with the gate insulator made of water-infiltrated nanoporous 12CaO$\cdot$7Al$_2$O$_3$ glass\cite{TE-FET1,TE-FET2}. On the other hand, an electric double layer transistor (EDLT) is one of FETs whose insulating layer is replaced with electrolyte and has an ability to accumulate carriers up to 10$^{14}$ $\sim$ 10$^{15}$ cm$^{-2}$ on the surface of the sample, which is one or two orders of magnitude larger than that in conventional FETs. This large amount of carrier accumulation lead to various achievement in condensed matter physics, such as metal-insulator transition in ZnO\cite{ZnOEDLT1,ZnOEDLT2}, field-induced superconductivity in SrTiO$_3$\cite{STOEDLT}, ZrNCl\cite{ZrNClEDLT}, KTaO$_3$\cite{KTOEDLT} and MoS$_2$\cite{MoS2EDLT1}, and formation of a p-n junction in MoS$_2$\cite{MoS2EDLT2}.
We have previously reported that the thermoelectric properties of ZnO have been controlled by using EDLT\cite{takayanagi}. In that report, the conductivity and thermopower have been modulated by gating and thus the power factor have been improved. The EDLT is therefore considered to be a powerful method to find new thermoelectric materials by controlling the carrier concentration.

In this paper, we propose new concept for thermoelectric device which can control thermoelectric properties of p-type and n-type materials simultaneously by electric double-layer gating. Here, the electric double-layer capacitor (EDLC) structure was applied to the thermoelectric device. The advantage of our technique is the simplification of design for the thermoelectric devices, compared to using FET structure. We do not need to use gate electrodes for each p-type and n-type materials. Moreover, EDLCs are already applied to practical use such as rechargeable battery. In order to design thermoelectric device, we used ZnO, which was already controlled thermoelectric properties by using EDLT, as a negative electrode. However, there are few materials, which shows typical p-type transistor behavior by electric double layer gating, up to date. NiO single crystal was reported as p-type material for EDLT channel with relatively small on/off ratio of 130\cite{NiOEDLT}. However, we could not reproduce EDLT device operation probably because we could not form electrodes of interdigitated comb shaped array which reduce effective device resistance. Several materials such as MoS$_2$\cite{MoS2EDLT2,MoS2EDLT3}, WSe$_2$\cite{WSe2EDLT} and Bi$_2$Te$_3$\cite{Bi2Te3EDLT} display both p-type and n-type conduction by changing a sign of gate voltage, but this ambipolar operation was confirmed only in the thin film or thin flake sample. In order to configure thermoelectric device, search of the p-type material with a good device performance in bulk sample as well as high thermoelectric properties is demanded. We investigated p-type material for EDLT, and found that Cu$_2$O which was grown by floating-zone (FZ) method displayed the typical p-type transistor behavior. Cu$_2$O is well-known as oxide semiconductor, which has a direct band gap of about 2.1 eV\cite{Cu2O}. Cu$_2$O usually shows p-type conductivity since Copper vacancies introduce an acceptor level $\sim$ 0.5 eV above the valence band. 
With using p-type Cu$_2$O and n-type ZnO as positive and negative electrodes, we have successfully accumulated the holes and electrons simultaneously on their surfaces by applying gate voltage between two electrodes. 

Single crystal of Cu$_2$O was grown by the FZ method\cite{Cu2OFZ}. A commercial Cu$_2$O (99.9 $\%$) powder was pressed into a rod and sintered at 1000 $\degC$ for 12 h in Pure Ar (99.99995 $\%$), so as not to be oxidized to CuO. The single crystal growth was carried out by using image furnace with two ellipsoidal mirror. The growth velocity was 3.5 mm/h and the growth atmosphere was air. The powder X-ray diffraction pattern for the crashed sample is good agreement with Cu$_2$O diffraction pattern. Hence, the grown crystal was confirmed to be Cu$_2$O.
ZnO polycrystalline samples were synthesized by sintering a commercial ZnO powder (99.99 $\%$) at 1000 $\degC$ for 15 h in air.
The samples were cut in rectangular shape (about 3$\times$1$\times$0.3 mm$^3$), and the surface was polished with an abrasive of 1 micron alumina lapping film to obtain the flat surface. 

Before performing the p-type and n-type simultaneous control, we evaluated the thermoelectric performance of each material by using EDLT configuration\cite{takayanagi}.
The schematic figure of EDLT is shown in Fig. 1(a). The source and drain contacts were made by Au paste, which was covered with silicone adhesive sealant to avoid a chemical reaction with the electrolyte. Pt foil was used as the gate electrode, and KClO$_4$ / Polyethylene glycol (PEG) was chosen as the electrolyte. The number-average molecular weight of PEG was 1,000 and the mixing ratio of KClO$_4$ / PEG was [K$^+$]:[O] = 1:20. The resistivity was measured by two-probe method. The absolute value of drain voltage ($V_D$) was 0.1V. In order to measure thermopower, a heater was attached near the source electrode as shown in Fig. 1(b). The temperature gradient $\Delta$T (approximately 1 $\sim$ 2K) was measured by a copper-constantan differential thermocouple and the sample voltage was measured between the source and drain electrodes. 

In order to control thermoelectric properties of p-type and n-type materials simultaneously, we applied the EDLC structure to the thermoelectric device as shown in Fig. 1(c)\cite{EDLC}. Compared to the EDLT configuration, the gate electrode of Pt foil was replaced to the p-type Cu$_2$O bulk sample. At 300 K, the PEG is gelatinous and ions can be easily moved by electric field. When the gate voltage ($V_G$) is applied between the p-type and n-type samples at 300 K, ions in the electrolyte are aligned on the surface of the samples, forming a charged double layer called Helmholtz layer. Therefore, a positive and negative image charges are induced on the surface of  Cu$_2$O and ZnO, respectively. 
However, the p-type and n-type materials must be connected in series for thermoelectric device as shown in Fig. 1(d). Since the PEG is perfectly frozen to lose ionic conductivity below 260K, Cu$_2$O and ZnO were connected below 260K, after applying $V_G$ at 300K. Accordingly, measurements of properties of whole device carried out at 250K. The device resistivity was measured by two-probe method with connecting p-type and n- type materials as shown in Fig. 1(d). The device thermopower was measured by attaining a heater to the connected side of the sample as shown in Fig. 1(e). 
Since the absorption of water into electrolyte causes electrochemical reaction on the interface of sample, all measurements were carried out under He atmosphere. 

Fig. 2 shows $V_G$ dependences of $I_D$, $I_G$ and $S$ of Cu$_2$O at room temperature, which was measured by EDLT configuration. The $V_G$ was swept from 0 V to -4 V, and then back to 0 V with the sweep rate of 0. 1V/min. The drain voltage $V_D$ was kept at 0.1 V during the measurement. No evidences of electrochemical reaction between the electrolyte and sample were observed, that is, no changes in $I_D$ and $S$ have been observed after sweeping $V_G$. $I_D$ increased by applying negative $V_G$ below -2 V, and level off at $V_G$ $\textless$ -3 V. This result confirms a typical p-type transistor behavior because the negative $V_G$ produces holes on the Cu$_2$O surface. The on-off ratio of the device was relatively small as 4, probably because Cu$_2$O sample used in this measurement had relatively low resistivity (7.5 k$\Omega$cm)\cite{ZnOEDLT2,takayanagi}. Therefore, the insulator-metal transition did not realized by electric double layer gating. The magnitude of leak current ($I_G$) was comparable to that in ZnO single crystal\cite{ZnOEDLT1,ZnOEDLT2} and ceramics\cite{takayanagi}, which confirm device quality. A slight hysteresis was observed in the forward and backward curves of $I_D$, implying the slow response of ions in the gelatinous electrolyte. As seen in Fig. 2(c), $S$ was clearly changed by using the EDLT configuration from 820 $\mu$V/K to 720 $\mu$V/K. 
This is the first result that the thermoelectric properties of p-type material are controlled by EDLT. 
Since the depth distribution of the accumulated carrier is not clear, it is difficult to estimate the resistivity and thermopower of the accumulated surface. However, if the accumulated surface and insulating bulk substrate are regarded as a parallel circuit for simplification, the observed total conductivity ($\sigma_{total}$) and thermopower ($S_{total}$) are given by 
\begin{eqnarray*}
\sigma_{total}=\sigma_{sur}+\sigma_{bulk},
\end{eqnarray*}
\begin{eqnarray*}
S_{total}=\frac{S_{sur}\sigma_{sur}+S_{bulk}\sigma_{bulk}}{\sigma_{sur}+\sigma_{bulk}},
\end{eqnarray*}

where $S_{sur}$ and $\sigma_{sur}$ are the thermopower and conductivity of the accumulated surface, and $S_{bulk}$ and $\sigma_{bulk}$ are those of the bulk substrate, respectively. As discussed in previous report, since the thickness of the surface accumulated layer is much smaller than that of the bulk substrate, we considered $\sigma_{bulk}$ as the conductivity at $V_G$ = 0 V. 

A large contribution from the bulk substrate exists in the conductivity in the whole range of gate voltage. While, only a slight decrease can be seen in the thermopower of the accumulated surface. These behaviors are due to the relatively large conductivity in bulk substrate and low on-off ratio.

By using the the same Cu$_2$O sample that used in the measurement of EDLT configuration, the simultaneous control of thermoelectric properties have been attempted with EDLC configuration. The $V_G$ was applied between the positive Cu$_2$O electrodes and the negative ZnO electrodes, in order to accumulate holes on the surface of Cu$_2$O and electrons on the surface of ZnO. First of all, the thermoelectric properties of each materials were measured separately, that is $I_D$ or electromotive force were monitored between source and drain contacts for each materials with changing $V_G$. 
Fig. 3 shows the $I_D$ and $S$ of p-type Cu$_2$O surface and n-type ZnO surface at 300K. 
Both materials showed a typical FET behavior, which indicate that we have successfully controlled the carrier concentration of p-type and n-type materials simultaneously.
The on-off ratio on the p-type channel was 3.8, which was in good agreement with Fig. 3(a). In contrast, the on-off ratio of n-type channel was approximately 20, which was slightly smaller than that reported in ref. [\citen{takayanagi}]. Since the bulk thermopower and resistance were smaller than the reported value, the carrier concentration of the sample used here may be slightly larger than that used in the previous report. The on-off ratio tends to decrease with decreasing resistivity because the off-state current is determined by the bulk resistivity. Thus, the on-off ratio of the device is likely to differ from sample to sample.
Considering these results, the accumulated carriers on the surface of the p-type and n-type materials are almost the same as that of the EDLT configuration.
In consequence, the controllability of the carrier concentration is considered to be comparable to the conventional EDLTs.

The gate voltage was removed after cooling the device to 250 K where the PEG was perfectly frozen and the accumulated carrier could not move. And then, the p-type and n-type materials were connected to measure the thermoelectric properties of the whole device.
The thermopower and resistance of the whole device ( $S_{dev}$ and $R_{dev}$, respectively) at 250K are shown in Fig. 4. $S_{dev}$ and $R_{dev}$ were obviously decreased by applying $V_G$. These results indicate that the carriers remain accumulated after cooling the device, although the p-type and n-type electrodes were connected. 
In the case of the thermoelectric device, $S_{dev}$ and $R_{dev}$ should be the sum of p-type and n-type materials, $S_{dev}$ = $S_p$ + $S_n$, $R_{dev}$ = $R_p$ + $R_n$, where $S_p$, $S_n$, $R_p$, $R_n$ are the thermopower and resistance of the p-type and n-type materials. The calculated $S_p$ + $S_n$ and $R_p$ + $R_n$ are also plotted in Fig. 4. As shown in Fig. 4, $S_{dev}$ and $R_{dev}$ were almost the sum of each materials. The increase of $S_{dev}$ and $R_{dev}$ are considered to be due to the temperature dependence of each materials, which increase with decreasing temperature. 

The output power of the device was estimated from $P$ = $S_{dev}^2$ / $R_{dev}$. The output power increased by applying $V_G$, where $P$ at $V_G$ = 4 V was more than 10 times larger than that at $V_G$ = 0 V. This indicate that the carrier concentration of the device can be optimize by electric double-layer gating. 

In conclusion, we have successfully controlled the thermoelectric properties of p-type and n-type materials simultaneously by electric double-layer gating, for the first time. In this study, EDLC configuration have been applied to the thermoelectric device, and the output power of the device have been increased to more than 10 times when the $V_G$ of 4V have been applied. Moreover, we confirmed that the controllability of the carrier concentration was comparable to that of the conventional EDLTs. 
Since the basic structure of our device is the same as EDLC which is already applied to practical use such as rechargeable battery, it has a potential ability for the practical application.

On the other hand, it was found that Cu$_2$O single crystal shows typical FET behavior with an on-off ratio of about 4. Since few materials have been reported as p-type material which shows typical FET behavior by electric double layer gating, further investigation of p-type materials will be required to make high performance thermoelectric device.

\newpage
\begin{figure}
\begin{center}
\includegraphics[width=12cm]{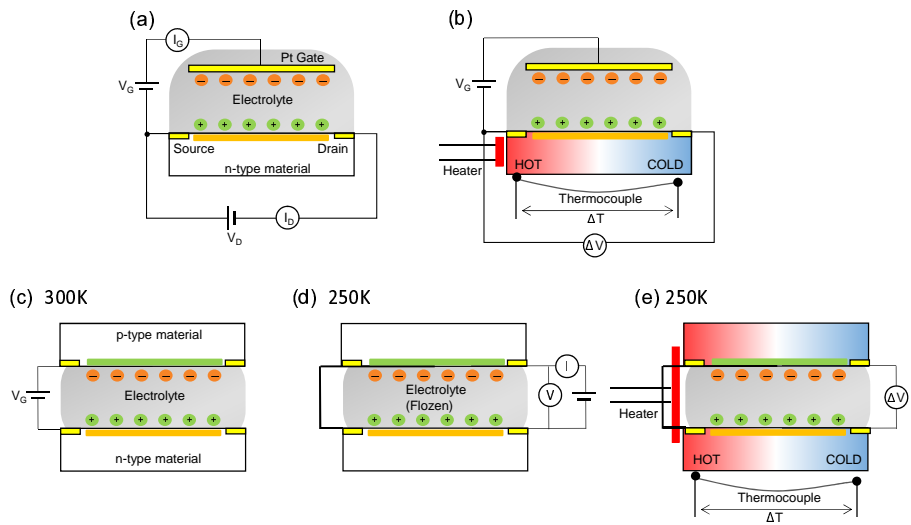}
\end{center}
\caption{(Color online) Schematic figures of electric double-layer transistor (EDLT) and electric double-layer capacitor (EDLC). (a) EDLT configuration for measuring resistance, and (b) that for measuring thermopower . (c) EDLC configuration (d) for measuring resistance, and (e) that for measuring thermopower. Fig.(a) and (b) are from ref. [\citen{takayanagi}]. "Copyright (2014) The Japan Society of Applied Physics."}
\end{figure}

\newpage
\begin{figure}
\begin{center}
\includegraphics[width=10cm]{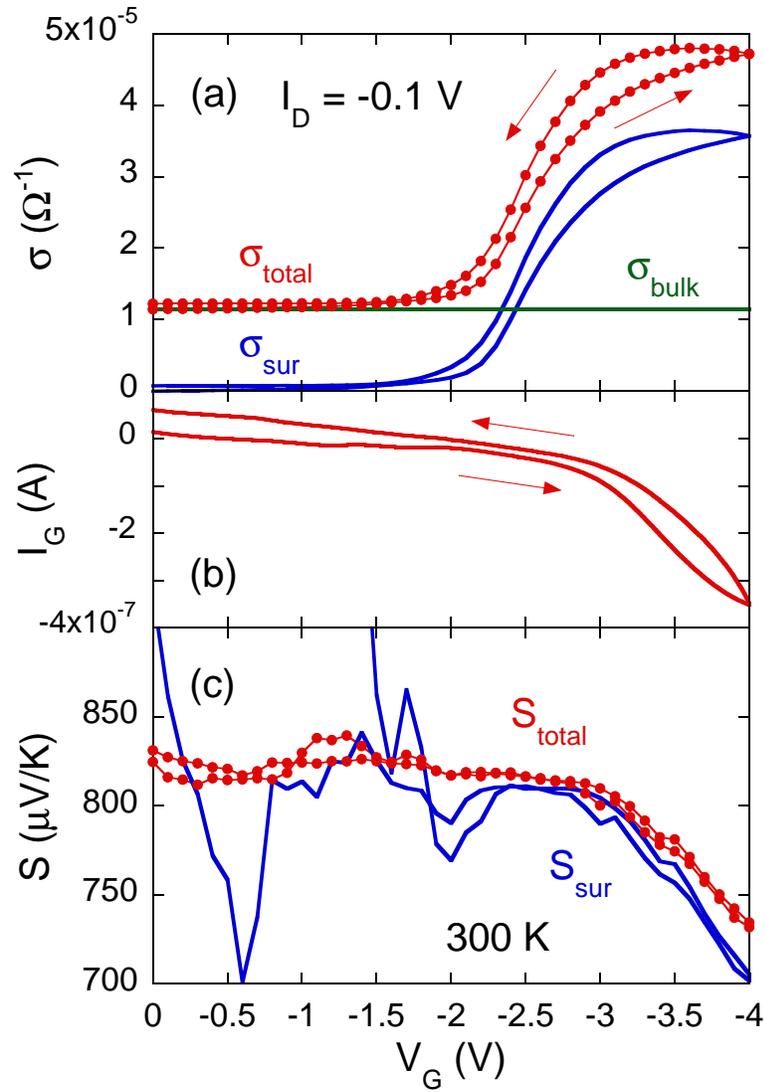}
\end{center}
\caption{(Color online) Transfer characteristics of Cu$_2$O at 300 K, (a) $I_D$-$V_G$, (b) $I_G$-$VG$ and (c) $S$-$V_G$ curve.}
\label{f1}
\end{figure}

\newpage
\begin{figure}
\begin{center}
\includegraphics[width=10cm]{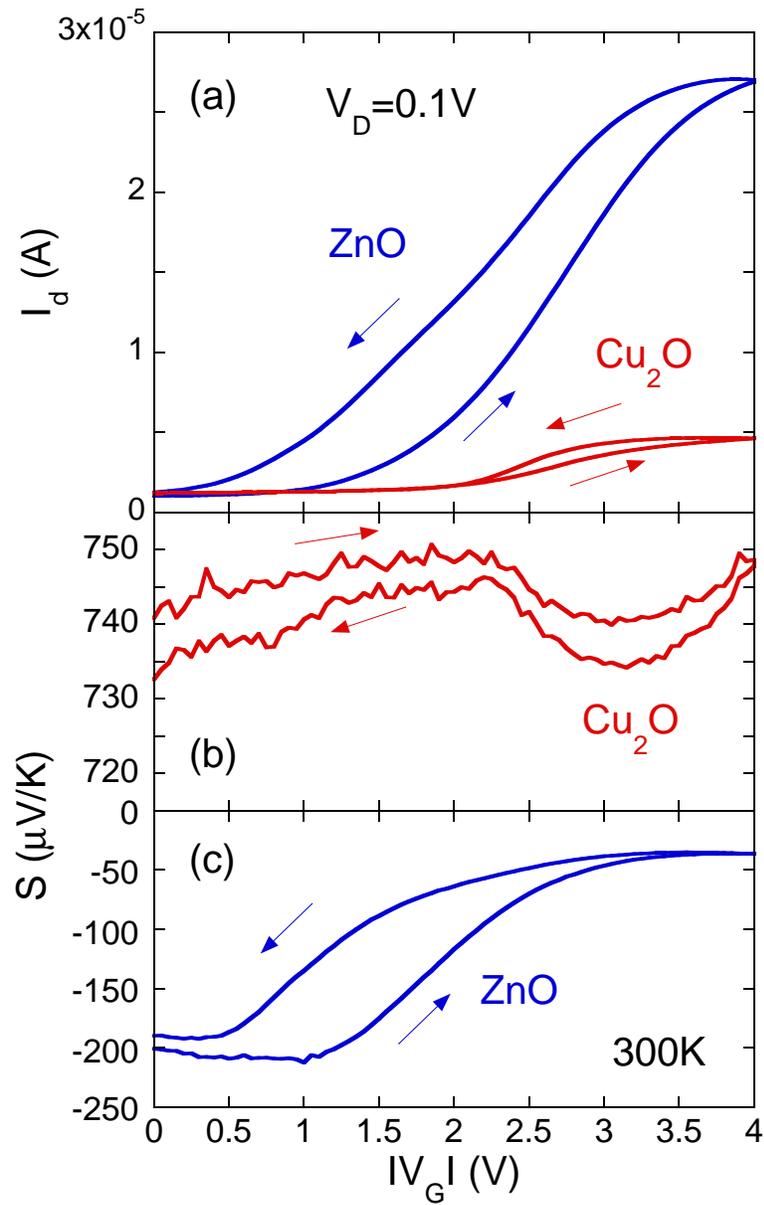}
\end{center}
\caption{(Color online) Transfer characteristics of Cu$_2$O (red line) and ZnO (blue line) with EDLC configuration at 300K: (a) $I_D$-$V_G$ of Cu$_2$O and ZnO (b) $S$-$V_G$ of Cu$_2$O (c) $S$-$V_G$ curve of ZnO.}
\label{f1}
\end{figure}

\newpage
\begin{figure}
\begin{center}
\includegraphics[width=10cm]{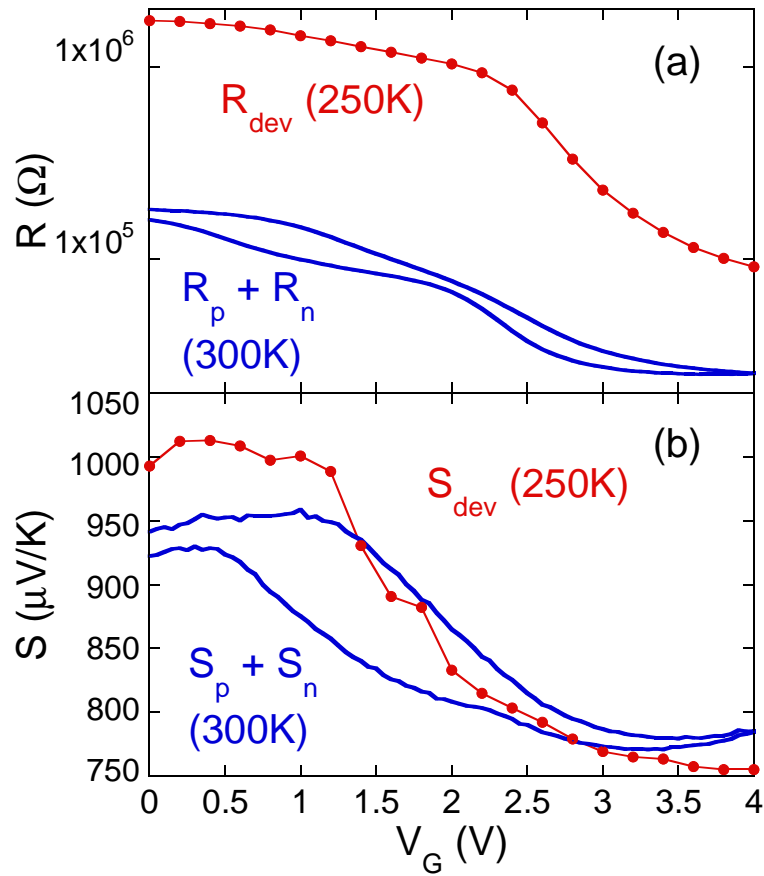}
\end{center}
\caption{(Color online) $V_G$ dependence of whole device at 250 K, (a) resistance and (b) thermopower. Red lines show the thermopower and resistance of the whole device. Blue lines show that of the sum of p-type and n-type materials}
\label{f1}
\end{figure}

\end{document}